\documentclass[twocolumn, amssymb,nobibnotes,aps,prb,nopacs]{revtex4}

\usepackage{amsmath}
\usepackage{graphicx}
\usepackage{subfig}
\usepackage[justification=RaggedRight,font=footnotesize]{caption}
\graphicspath{ {figs/} }

\usepackage{xcolor}

\usepackage{multirow}

\begin{document}

\title{Scattering of electron holes in the context of ion-acoustic regime}

\author{S. M. Hosseini Jenab\footnote{Email: Mehdi.Jenab@umu.se}}
\author{F. Spanier \footnote{Email: Felix@fspanier.de}}
\author{G. Brodin  \footnote{Email: Gert.Brodin@umu.se}}
\affiliation{Department of Physics, Faculty of Science and Technology, Ume\aa \ University, 90187, Ume\aa , Sweden}
\affiliation{Centre for Space Research, North-West University, Potchefstroom Campus,
Private Bag X6001, 2520, Potchefstroom, South Africa}

\date{\today}

\begin{abstract}
Mutual collisions between ion-acoustic (IA) solitary waves are studied based on a fully kinetic simulation approach.
Two cases, small and large relative velocity,  are studied and the effect of trapped electron population 
on the collision process are focused upon. 
It is shown that, for the case of small relative velocity, 
the repelling force between the trapped populations of electrons results in scattering of electron holes.
However, this phenomenon can not be witnessed if the relative velocity is considerably high,
since the impact of trapped population stays very weak.
\end{abstract}
\maketitle
Solitary waves, nonlinear localized structures,
can stay unaltered for a long propagation\cite{Wadati2001841}. 
When mutual collisions take place between them, 
two profiles start overlapping and merge. 
Some time, their profiles reappears after the collision (overlapping stops) 
with no alteration in their features, i.e.  velocity, height, width and 
shape in both velocity and spatial directions.
In the context of plasma physics, the discovery of ion-acoustic (IA) solitons\cite{Zabusky1965}
sparked a long-lasting fascination with this concept\cite{Scott2007nonlinear, tran1979ion,shafranov2012reviews}.
Most studies, both theoretical and simulation,
employ fluid approach.
These simulations - either KdV or full fluid- ignore the kinetic effects \cite{Kakad2013, kakad2014nonlinear, Sharma2015}.
It is shown that for large amplitude solitons even the full-fluid simulations
strongly diverge from kinetic simulations \cite{kakad2014nonlinear}.
The importance of solitary waves to experimental observations in space or the laboratory
has been discussed in details in some review papers and previous papers of authors\cite{jenab2016trapping,deng2006observations,catte1151998fast,hobara2008cluster,pickett2004solitary,kuznetsov1986soliton}.

Nonetheless, theoretical approaches such as the Sagdeev pseudo-potential\cite{Sagdeev} 
and the BGK method\cite{bernstein1957exact} provide a platform to study
IA solitary waves on kinetic level.
Although they can not predict the temporal evolution, 
they are able to provide the static shape of solitary waves\cite{schamel_2,schamel_3}.
Due to the lack of temporal aspects in these theories,
not only does the propagation of solitary waves stay beyond their scope, 
but also their collisions can not be studied.
On the kinetic level, i.e. considering distribution functions, 
IA solitary waves trap electrons. 
The trapped population appears as vortex-shape structures 
in the phase space and as a hump in the density profile.
Schamel suggested a distribution function to model these vortex shapes structures\cite{schamel_1}.

The concept of "soliton" is used with different meanings in the
literature. A localized pulse that moves with fixed shape, due to a
nonlinearity that counteracts dispersion, is normally called a solitary
wave. In case solitary waves survive collisions with each other,
preserving speed, amplitude and pulse shape, one sometimes label these
solitary waves as solitons. However, in a strict sense, solitons should
satisfy a number of mathematical criteria, in particular obeying an
exactly integrable nonlinear evolution equation, and having an infinite
number of conserved quantities. The latter is for example satisfied by
the wellknown examples of KdV-solitons and NLS-solitons.  In the present
paper we will perform a numerical study, and hence we can only establish
soliton type of behavior  in a looser sense. As a result we will refer
to "soliton-like stuctures", whenever solitory waves preserve their
identities during collisions, in order to distinguish from the concept
of solitons in a more strict mathematical meaning.

In recent attempts, mutual collisions of  IA soliton-like structures are reported on the kinetic level. 
It is shown that the kinetic effects, mainly electron trapping, 
causing a more complicated behavior in the distribution function during mutual collision
compared to fluid level\cite{jenab2017pre,jenab2017overtaking}.
There has been studies on the collision process of electron holes and the effects of ion motion on them
\cite{mandal2016nonlinearly,saeki1998electron, eliasson2006formation}. 
Based on a fully kinetic simulation approach, it is shown that  IA solitary waves with trapped electrons survive mutual collisions, 
either head-on \cite{jenab2017pre} or overtaking \cite{jenab2017overtaking}.
Hence we can refer to the solitary waves as soliton-like structures.

Here, the focus is on the influence of trapped electrons and their relative velocity on the collision process. 
Since these trapped population act as pseudo-particles carrying their own charges, 
their repelling electrostatic interaction can cause the electron holes to slow down when approaching each other \cite{dupree1983growth}. 
In the set of simulations presented here, by choosing the electron holes with close propagation velocities, 
we are able to show these subtle effect. 
Initially, the case of large relative velocity is presented which serves as benchmarking test of the simulation code.
In the main part of the paper, the case of small relative velocity is shown which presents the 
scattering of electron holes from each other.

Note that our focus in this study is on the positive potential profiles which traps electrons. 
The effect of ions trapping on the negative potential profiles has been discussed in references such as \cite{schamel2018privileged, omura1996electron}.

\begin{table}
\small
\caption{Normalization of quantities.}
\begin{ruledtabular}
\begin{tabular}{cccc}
   \multirow{2}{*}{Name}&   \multirow{2}{*}{Symbol}& 
   \multicolumn{2}{c}{Normalized by}  \\
  {}&  {}&  Name& formula \\
 \hline
  Time     & $\tau$  	&ion plasma frequency 		&$\omega_{pi}  = {\big(\frac{n_{i0} e^2}{m_i \epsilon_0}\big)^{\frac{1}{2}} }$   \\
  Length   & $L$  	&ion Debye length		&$\lambda_{Di} = \sqrt{ \frac{\epsilon_0 K_B T_i}{n_{i0} e^2}   }$   \\
  Velocity & $v$  	&ion thermal velocity		&$v_{th_i} = \sqrt{\frac{K_B T_i}{m_i}}$   \\
  Energy   & $E$  	&{ion thermal energy}			&$K_B T_i$   \\
  Potential   & $\phi$  	&{-------}		&$\frac{K_B T_i}{e}$   \\
  Charge 		&$q$		&elementary charge &$e$ \\
  Mass 			&$m$		&ion mass		&$m_i$
\end{tabular}
\end{ruledtabular}
\label{table_normalization}
\end{table} 
The normalization of equations and quantities are based on the table \ref{table_normalization}.
Hence, the scaled set of equations read as follow:
\begingroup\makeatletter\def\f@size{9}\check@mathfonts
\def\maketag@@@#1{\hbox{\m@th\large\normalfont#1}}%
\begin{align*}
\frac{\partial f_s(x,v,t)}{\partial t} 
+ v \frac{\partial f_s(x,v,t)}{\partial x} 
&+  \frac{q_s}{m_s} E(x,t) \frac{\partial f_s(x,v,t)}{\partial v} 
= 0, \\
\frac{\partial^2 \phi(x,t)}{\partial x^2} & = n_e(x,t) - n_i(x,t)
\label{Vlasov}
\end{align*}\endgroup
where $s = i,e$ represents the corresponding species, i.e. electrons and ions.
They are coupled by density integrations for each species to form a closed set of equations:
\begingroup\makeatletter\def\f@size{9}\check@mathfonts
\def\maketag@@@#1{\hbox{\m@th\large\normalfont#1}}%
\begin{align*}
n_s(x,t) = n_{0s} N_s(x,t), \ \ 
N_s(x,t) = \int f_s(x,v,t) dv
\end{align*}\endgroup
in which $N$ stands for the number density.

The Schamel distribution function \cite{schamel_1} has been utilized
as the initial distribution function:
\begingroup\makeatletter\def\f@size{8.3}\check@mathfonts
\def\maketag@@@#1{\hbox{\m@th\large\normalfont#1}}%
\begin{equation*}
f_{s}(v) =  
  \left\{\begin{array}{lr}
     A \ exp \Big[- \big(\sqrt{\frac{\xi_s}{2}} v_0 + \sqrt{\varepsilon(v)} \big)^2 \Big]   &\textrm{if}
      \left\{\begin{array}{lr}
      v<v_0 - \sqrt{\frac{2\varepsilon_{\phi}}{m_s}}\\
      v>v_0+\sqrt{\frac{2\varepsilon_{\phi}}{m_s}} 
      \end{array}\right. \\
     A \ exp \Big[- \big(\frac{\xi_s}{2} v_0^2 + \beta_s \varepsilon(v) \big) \Big] &\textrm{if}  
     \left\{\begin{array}{lr}
      v>v_0-\sqrt{\frac{2\varepsilon_{\phi}}{m_s}} \\
      v<v_0 + \sqrt{\frac{2\varepsilon_{\phi}}{m_s}} 
      \end{array}\right.
\end{array}\right.
\label{Schamel_Dif}
\end{equation*}\endgroup
in which $A = \sqrt{ \frac{\xi_s}{2 \pi}} n_{0s}$,
and $\xi_s = \frac{m_s}{T_s}$ are the amplitude and the normalization factor respectively.
$\varepsilon(v) = \frac{\xi_s}{2}(v-v_0)^2 + \phi\frac{q_s}{T_s}$ 
represents the (normalized) energy of particles in which $\varepsilon_{\phi} = q_s \phi$.
This invokes a self-consistent localized compressional profile in density, in other words initial density perturbation (IDP).
The IDP is characterized by a nonlinear structure in the phase space of electrons which can take three different forms, namely 
hollows, humps and plateau based on a variable called \textit{trapping parameter} ($\beta$).
We start with an IDP at rest ($v_0 = 0$), that breaks into to two oppositely drifting density perturbations (DDPs). 
Each of the DDPs, later on, breaks into a number of IA solitary waves, hence the IA solitary waves are produced self-consistently in
the following simulations \cite{jenab2016trapping}. 
The velocity of solitary waves stays slightly above the ion-acoustic speed\cite{jenab2016trapping}.

On next step, these IA solitary waves are isolated and arranged in a periodically bounded simulation boxes to create different
scenarios of collisions.
Note that trapped population of the electrons accompanying the IA solitary waves possess the same 
trapping parameter ($\beta$), hence the same form in the phase space,
as the IDP that they are originated from. 
Simulations are one dimensional (1D+1V) and the ratio of electron and ion mass and temperature are $\frac{m_i}{m_e} = 100$ and $\frac{T_e}{T_i} = 64$ respectively.
For more details see Ref. 7\nocite{jenab2016trapping}. 

The length of the simulation box is $L_x = 1024$ and $(v_{min}, v_{max}) = (-600, 600)$ for electrons.
For ions the velocity box is $(v_{min}, v_{max}) = (-6, 6)$.
The grid size for both species is $(N_x, N_v) = (1024,4000)$.
Although the simulations are carried out for different values of $\beta$,
here we are just presenting the results with $\beta = -0.1, 0.0, 0.2$ which will be specified accordingly.


\begin{figure}
  \subfloat{\includegraphics[width=0.5\textwidth]{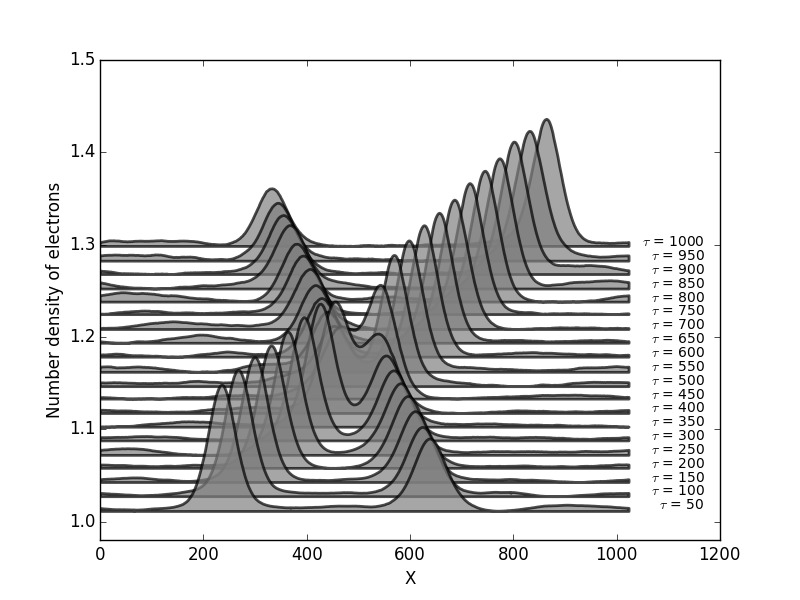}} \\
  \subfloat{\includegraphics[width=0.5\textwidth]{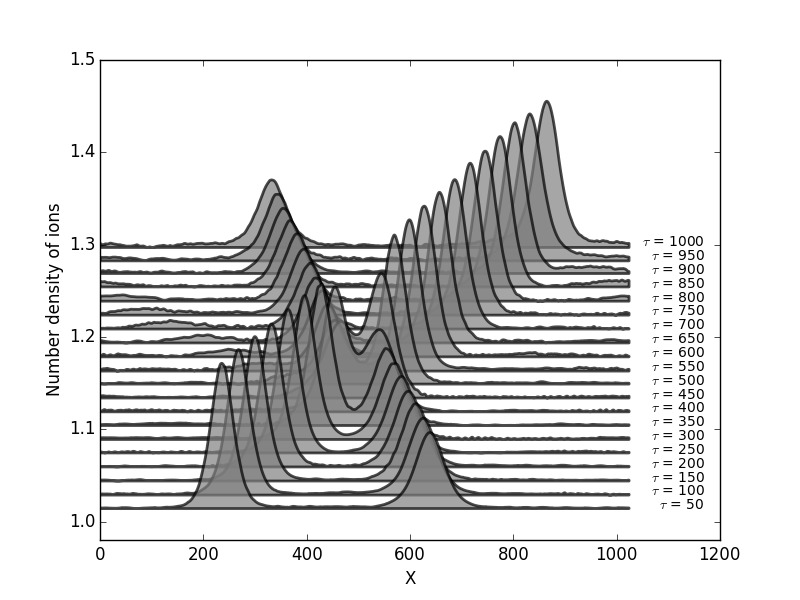}}
  \caption{Overtaking collision of two IA solitary waves is presented in temporal evolution of number density 
  of (a) electrons and (b) ions.
  Starting from time $\tau = 50$ (the lowest line) to $\tau = 1000$ (the highest line) shown in $20$ intervals.}
  \label{Fig_multiple_OT}
\end{figure}

\begin{figure*}
  \subfloat[]{\includegraphics[width=0.3\textwidth]{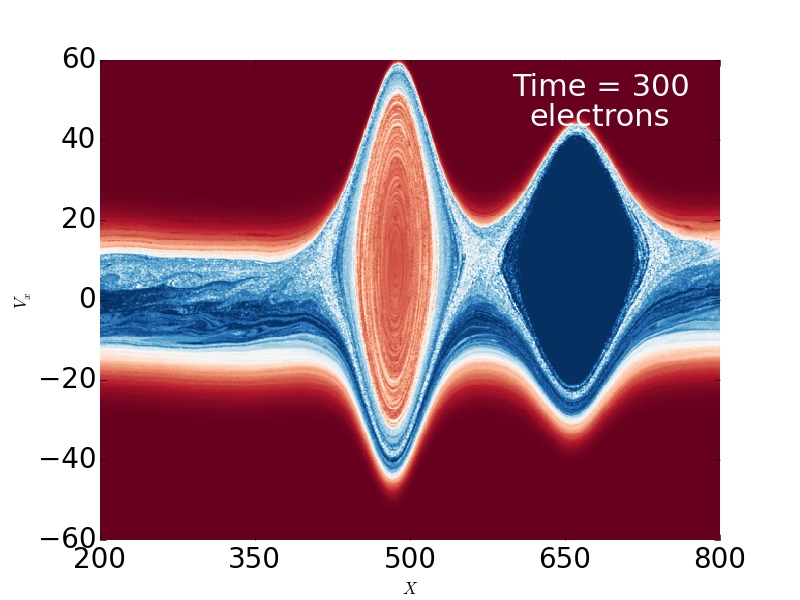}}
  \subfloat[]{\includegraphics[width=0.3\textwidth]{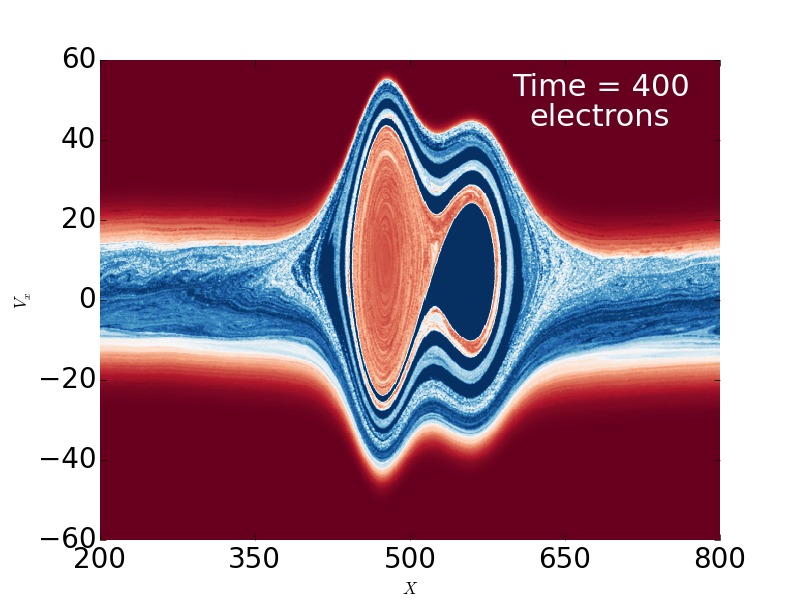}}
  \subfloat[]{\includegraphics[width=0.3\textwidth]{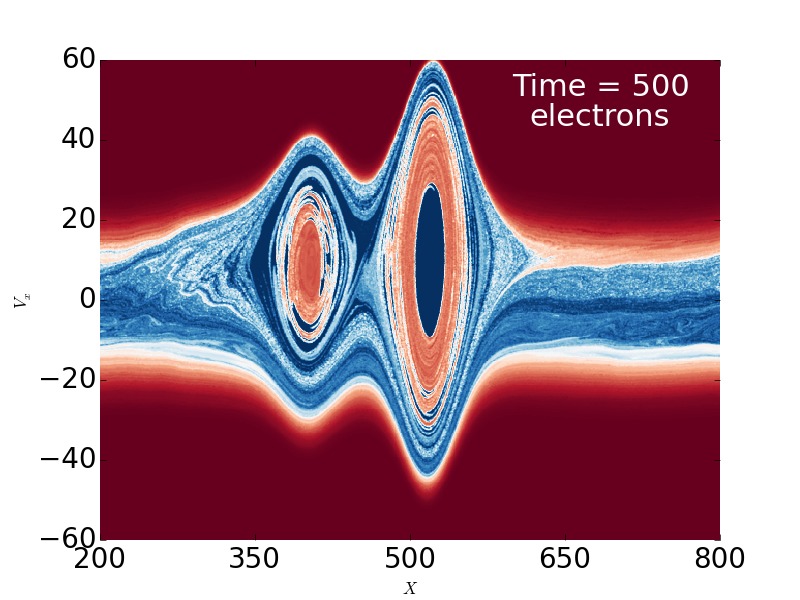}}
  \subfloat{\includegraphics[width=0.085\textwidth]{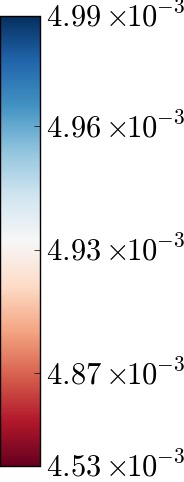}}
  \caption{Phase space of electrons are shown in three different time steps in case of overtaking collision, namely
  before ($\tau = 300$), during ($\tau = 400$) and after ($\tau = 500$) the collision.
  Before the collision ($\tau = 300$), solitary waves on the left (right) is accompanied by a hole (plateau) with $\beta = -0.1$ ($\beta = 0$).}
  \label{Fig_Overtaking_DF}
\end{figure*}

Fig.~\ref{Fig_multiple_OT} displays the results for a collision between two IA solitary waves
with a considerable difference between their velocities. 
For such cases, the small/secondary effects of repulsion, originating from the same charges of trapped populations, 
can not impose any noticeable influence on the collision process. 
Therefore, IA solitary waves come close to each other, their profiles merge and afterwards they 
reappear without any change in their features except for the small phase shift in their trajectories, 
i.e. soliton-like behavior.

Fig. \ref{Fig_Overtaking_DF} displays the phase space structures for the aforementioned collision, 
in the electrons distribution function. 
The right (left) propagating solitary waves is accompanied by a hole (plateau) before the collision (see Fig. \ref{Fig_Overtaking_DF}.a)
with $\beta=0$ ($\beta = -0.1$).
When they overlap, exchange of trapped population happens (Fig. \ref{Fig_Overtaking_DF}.b). 
Finally, they depart in their direction as before the collision but with some portion of the trapped population 
acquired from each other.
Note that, despite all the alteration in the internal structure of the trapped population in the phase space, 
their features on the fluid level (number density profiles) stays the same as before the collision.
Furthermore, on the kinetic level, the overall shape and area of the trapped population goes unaltered as well.
For more details discussion on this see Refs. 5 and 6 \nocite{jenab2017overtaking,jenab2017pre} 
in which the stability versus mutual collisions
is extensively discussed.
Collision process through kinetic theoretical treatment is also discussed in Ref.[28]\nocite{zhou2018dynamics}.

\begin{figure} [!h]
  \subfloat{\includegraphics[width=0.5\textwidth]{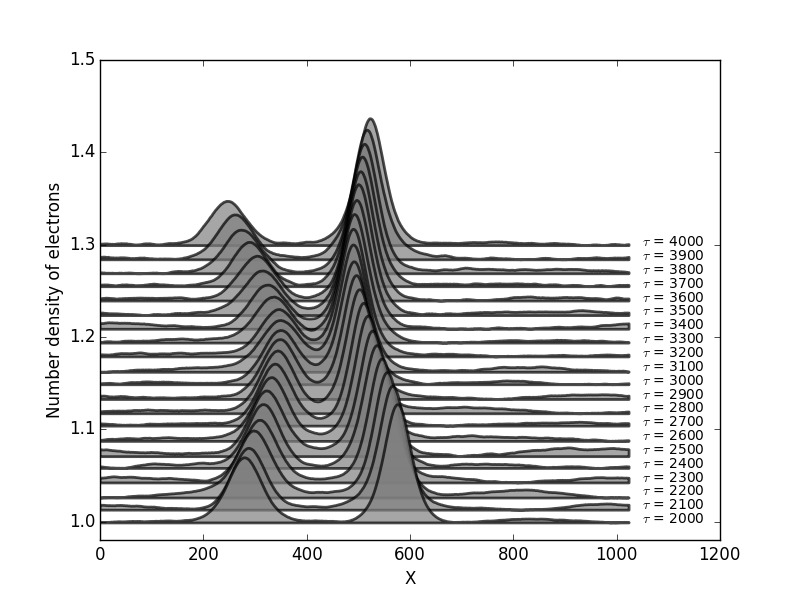}} \\
  \subfloat{\includegraphics[width=0.5\textwidth]{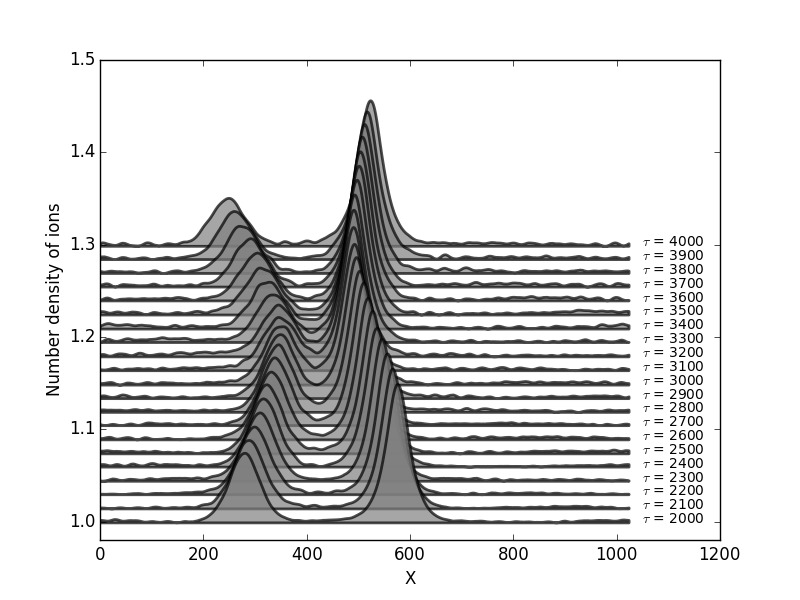}}
  \caption{Scattering of two IA solitary waves is shown in temporal evolution of number density of (a) electrons and (b) ions.
  Starting from time $\tau = 2000$ (the lowest line) to $\tau = 4000$ (the highest line) shown in $20$ intervals.}
  \label{Fig_multiple_SC}
\end{figure}

Fig. \ref{Fig_multiple_SC} shows the results of simulation for a mutual collision
between two IA solitary waves with comparatively small relative velocity. 
The collision process exhibits a remarkable difference with the previous case studied
in Figs. \ref{Fig_multiple_OT} and \ref{Fig_Overtaking_DF}.
The two IA solitary waves approach each other, they start the overlapping process. 
However, this stops and they bounce off each other continue their propagation in the opposite direction. 
Note that the profiles here are shown in the window moving with their relative velocity.

\begin{figure*}
  \subfloat{\includegraphics[width=0.35\textwidth]{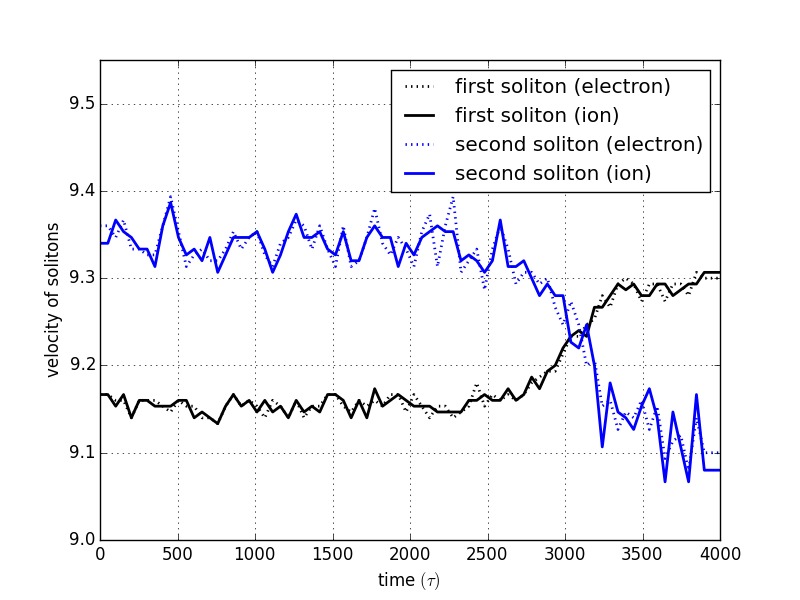}}
   \subfloat{\includegraphics[width=0.35\textwidth]{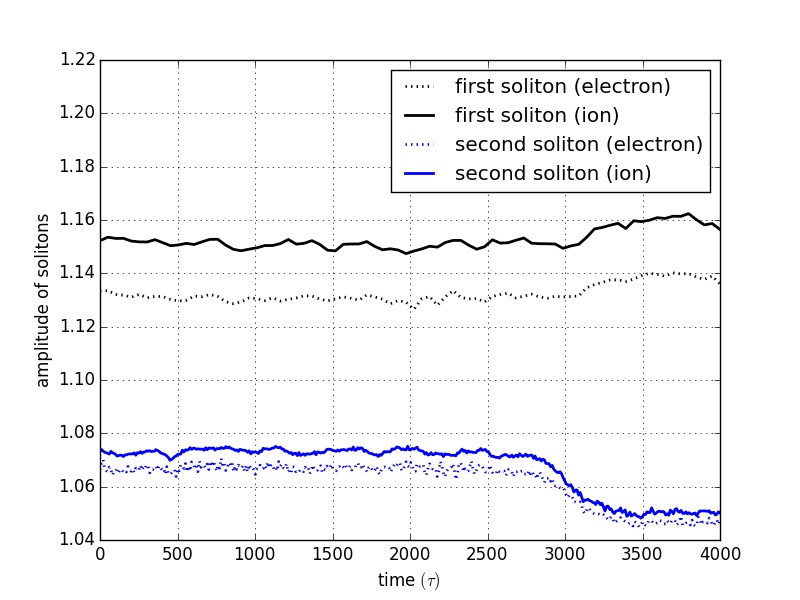}}
   \subfloat{\includegraphics[width=0.35\textwidth]{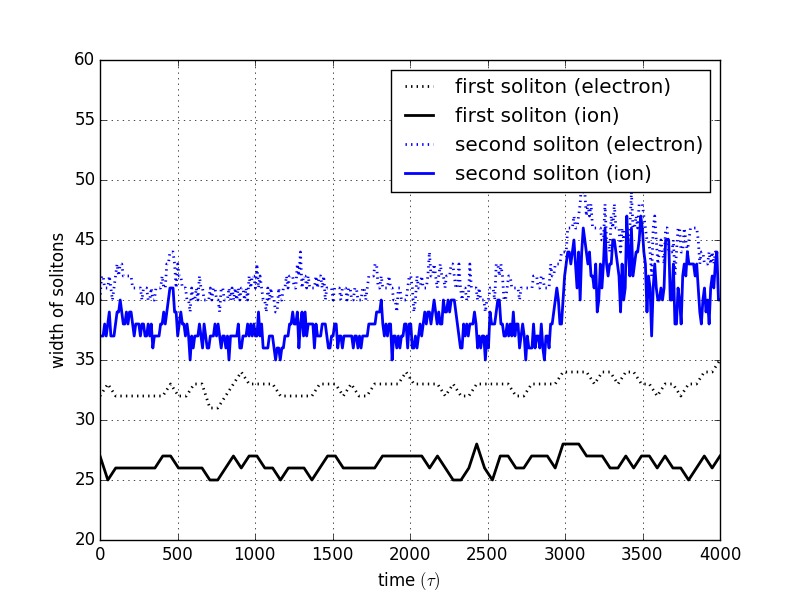}}
  \caption{Features of number densities of plasma components, i.e. electrons and ions, are presented for 
  the two electron holes involved in scattering process. The scattering causes them to change velocity with each other.
  However, the other characteristics, i.e. width and amplitude, stay the same.}
  \label{Fig_Scattering_features}
\end{figure*}

The analysis of the two solitary waves' features, both for electron and ion profiles, are shown in Fig.~\ref{Fig_Scattering_features}.
The change in their velocities can be recognized which is a symmetric exchange of velocity between them. 
Since two solitary waves carry roughly the same number of trapped particles, they possess almost equal mass and the
conservation of momentums dictates this symmetry.
There exist some small changes in the amplitude and width, which take place due to the short overlapping and the exchange of 
trapped population during that time. 

\begin{figure*}
\begin{tabular}{cccc} 
  \subfloat{\includegraphics[width=0.3\textwidth]{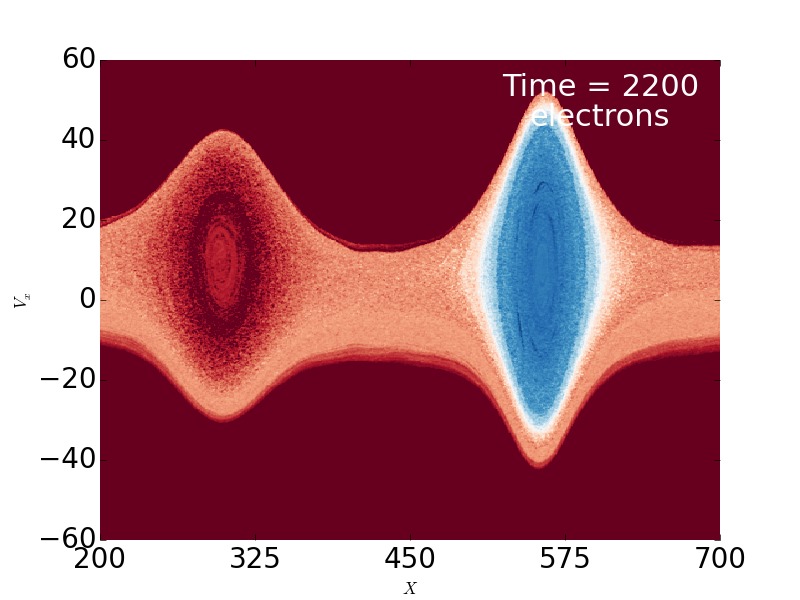}}&
  \subfloat{\includegraphics[width=0.3\textwidth]{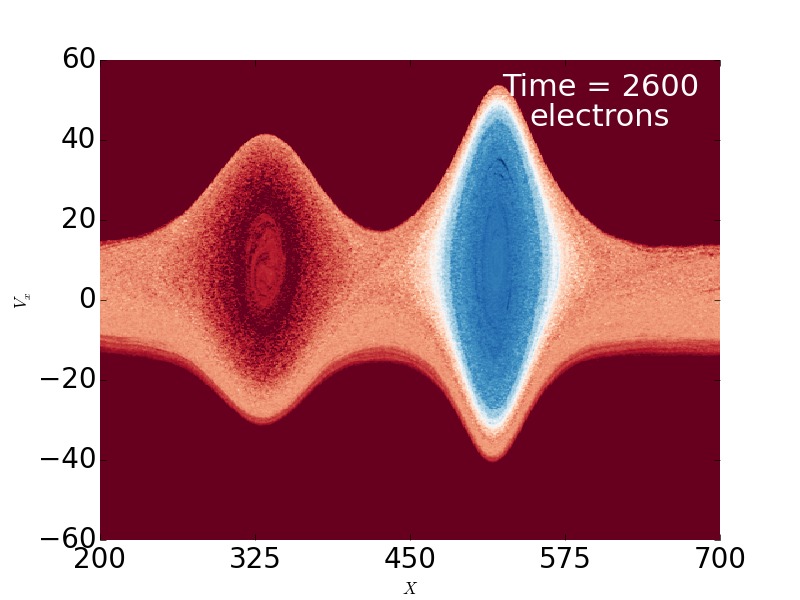}} &
  \subfloat{\includegraphics[width=0.3\textwidth]{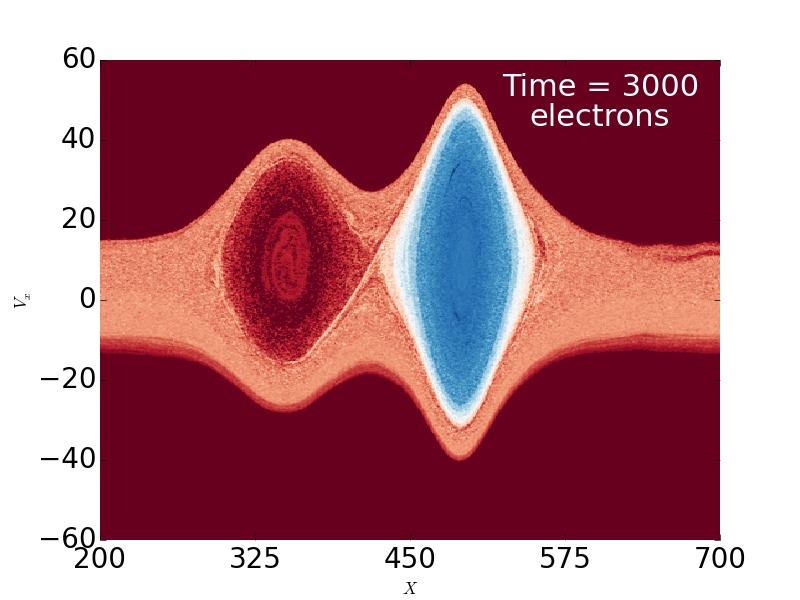}} &
   \multirow{-3}[18]{*}{{\includegraphics[width=0.07\textwidth]{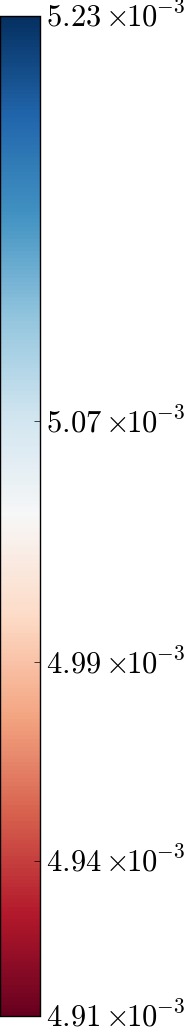}}}  \\ 
  \subfloat{\includegraphics[width=0.3\textwidth]{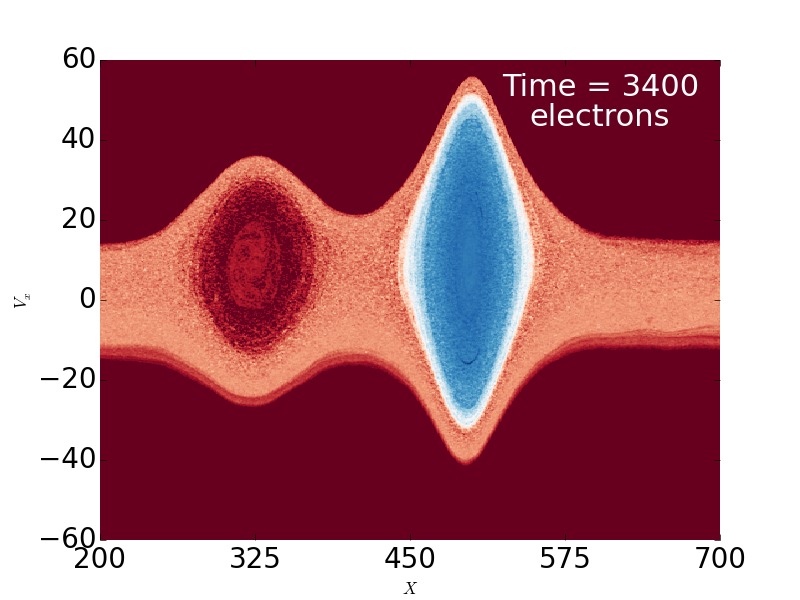}}&
  \subfloat{\includegraphics[width=0.3\textwidth]{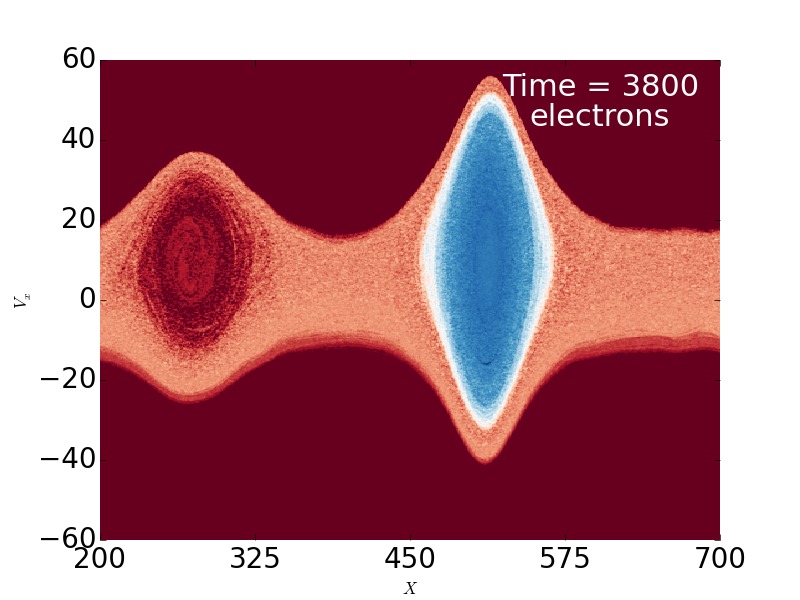}}&
  \subfloat{\includegraphics[width=0.3\textwidth]{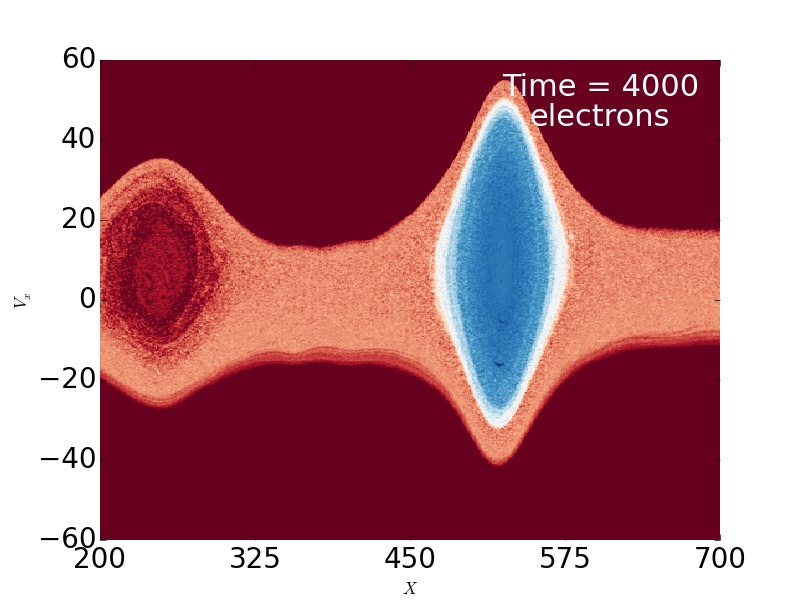}}&
  {}
\end{tabular}
  \caption{scattering of two electron holes which is shown in temporal evolution of their associated trapped population of electrons
  in the phase space. The left (right) structure is accompanied by a hollow (hump) with $\beta = 0$ ($\beta = 0.2$).}
  \label{Fig_Scattering_DF}
\end{figure*}

Furthermore, the details of temporal progression of electrons distribution function
of trapped population are presented Fig. \ref{Fig_Scattering_DF} during collision.
The overlapping takes place at $\tau = 3000$, and afterwards, the two depart in the opposite direction.
Comparing this behavior with the collision of large relative velocity highlights 
(see Figs. \ref{Fig_Overtaking_DF} and \ref{Fig_multiple_SC}) 
the sharp different between temporal evolution of the two cases.
Our interpretation of the process suggests that electrostatic forces between 
the two trapped populations of electrons, which are carrying the same charge, causes them to repel each other.
Consequently, this causes the two electron holes to experience scattering phenomena. 

We have carried some more simulations to determine the effect of trapping parameter $\beta$ on the 
process of the scattering. 
Our results show that the effect of trapping parameter is negligible if the overall shape and 
amplitude of the nonlinear solutions stay the same.

In summary, we have shown the existence of electron holes scattering from each other for ion-acoustic (IA) regime 
based on a fully kinetic simulation approach. 
By adopting the chain formation process, we have managed to create self-consistent
IA solitary waves from an initial density perturbation (IDP)\cite{jenab2016trapping}. 
By using them in different scenarios, we have shown
that when the relative velocity of two electron holes are small enough,
their trapped population of electrons with the same charge 
play the role of repelling force and cause them to bounce off each other. 

G. Brodin and S. M. Hosseini Jenab would like to acknowledgment financial support 
by the Swedish Research Council, grant number 2016-03806.
FS likes to thank the Deutsche Forschungsgemeinschaft (DFG) for support through grant SP 1124/9.
This work is based upon research supported by the National Research Foundation 
and Department of Science and Technology.
Any opinion, findings and conclusions or recommendations expressed in this 
material are those of the authors and therefore the NRF and DST do not accept 
any liability in regard thereto.

%


\end{document}